# Cosmic Strings in an Open Universe with Baryonic and Non-Baryonic Dark Matter

Pedro Ferreira

*Blackett Laboratory, Imperial College, South Kensington, London SW7 2BZ, U.K.*

(October 11, 1994)

## Abstract

We study the effects of cosmic strings on structure formation in open universes. We calculate the power spectrum of density perturbations for two class of models: one in which all the dark matter is non baryonic (CDM) and one in which it is all baryonic (BDM). Our results are compared to the 1 in 6 IRAS QDOT power spectrum. The best candidates are then used to estimate $\mu$, the energy per unit length of the string network. Some comments are made on mechanisms by which structures are formed in the two theories.

Typeset using REVTeX

The cosmic string scenario has generated much interest over the past few years as a promising candidate for structure formation. Following a primordial phase transition, line-like concentrations of energy could form in certain Grand Unified Theories [1]. These would then evolve into a scaling regime, in which all correlation functions should be expressed in terms of dimensionless ratios of length scales over the horizon. The long-string energy density would evolve as

$$\rho_L \propto \frac{1}{t^2} \qquad (1)$$

perturbing the background radiation and matter density, leading to the formation of astrophysical objects and anisotropies in the microwave background.

Albrecht & Stebbins have performed an analysis of string-seeded perturbations in non-baryonic Cold Dark Matter (CDM) and Hot Dark Matter (HDM) [2]. They have developed a model (form here on the AS model) which not only allows for different possibilities for string network evolution and matter content of the universe but also includes the effect of compensation. They have found that, in a CDM universe, strings produce too much power on small scales relative to large scales, more so than in an adiabatic CDM cosmology. They propose however that in a HDM cosmology the freestreaming by the neutrinos will partially suppress perturbations on small scales to a level which might be consistent with the data. Recent analysis of large scale surveys have indicated that the spectrum of mass fluctuations tends to flatten out on scales larger than in standard ($\Omega = 1$) CDM or HDM models, and consequently the effect of HDM on string perturbations would not be effective on large enough scales. Thus strings and HDM seems to have some of the same problems as its CDM counterpart.

Although there are strong theoretical reasons to favour an $\Omega = 1$ universe, as yet there is no strong observational evidence for it (see [4]). On the other hand, the extension of standard cosmological scenarios to open universes has proven to be quite promising; CDM adiabatic and topological defect models do not have excess small scale power if one resorts to low $\Omega h$ universes. This renewed interest in open universes has led us to reanalize the AS model. Due to the large number of choices we can make (string network model, type and density of dark matter, $H_0$, ionization history, etc) we have opted to present the results of what we think are representative models. We will consider standard recombination, $H_0 = 50 \mathrm{Km s^{-1} Mpc^{-1}}$, universes where all the dark matter is either non-baryonic or baryonic (BDM) and we have chosen one model for the network evolution. In a future publication we shall present a more detailed description of parameter space.

We shall briefly illustrate the formalism we are using. One is interested in calculating the power spectrum of density perturbations, defined as $(2\pi)^3 P(k) \delta^{(3)}(\mathbf{k} - \mathbf{k}') = \langle \delta(\mathbf{k}) \delta(-\mathbf{k}') \rangle$. The AS model uses the form

$$P(k) \simeq 16\pi^2 (G\mu)^2 \int_{\eta_i}^{\eta_0} |G_M(k, \eta_0, \eta')|^2 \mathcal{F}(k\xi/a) d\eta' \qquad (2)$$

where:

1) The "structure function": as proposed in [2],

$$\mathcal{F} = \frac{2}{\pi^2} \beta^2 \Sigma \frac{\chi^2}{\xi^2} \left( \frac{1}{1 + 2(k\chi/a)^2} \right) \qquad (3)$$

is the time integral of the two-point correlation function of the "$\rho + 3P$" part of the long string stress energy tensor. We must emphazise that it is specific to cosmic string physics; it has a $k^{-2}$ behaviour on scales smaller than the long string coherence scale, characteristic of sheets, and a $k^0$ behaviour on large scales, typical of uncorrelated point-like objects. The different parameters are $\xi = (\rho_L/\mu)^{\frac{1}{2}}$, the long string coherence or "curvature" scale $\chi$, the r.m.s. velocity of strings $\beta$ and the amount of small scale structure $\Sigma$.

2) The string network: we are assuming that the main source of perturbations are the long strings [5]. There are no numerical simulations of string networks in a curvature dominated universe and all current efforts to find reliable analytical models have focused on flat universes. We have modeled the string network in three different ways. Having determined the differences to be unimportant we have chosen one model for subsequent calculations. Model 1 is the original idealized model which has the correlation length scaling with the horizon, the X model in [2]. Model 2 is a good fit to the flat-spacetime simulations and is given in [6]; we extrapolate to an open universe using the parameters calculated deep in the radiation era in [7] [8]. In this case, when curvature starts to dominate, strings will slow down and one finds that the network will deviate from scaling. We find the same sort of behaviour in model 3 proposed in [3], motivated by large-N non-linear sigma model dynamics and by nematic liquid crystals where evolution is friction dominated. Here $\xi \simeq \eta^{\frac{1}{2}}$ in the free-expansion phase, so one can interpolate between the scaling and nonscaling behaviour with $\chi = \beta(a/\mathcal{H})^{\frac{1}{2}}$. One finds that, on the scales we are interested, deviation from scaling will have no significant effect. In fact all these models give very similar powerspectra albeit with different normalizations. We have chosen to work with model 2 so $\chi = 2\xi$, $\Sigma\beta^2 = 1.2$, and $k_c = 4\pi/\eta$.

3) Green's function: $G_M$ is the Green's function of the matter density perturbation which undergoes gravitational collapse at late time. The equations are solved in the synchronous gauge (all perturbations are seeded inside the horizon so we don't have to deal with gauge issues) and $k$ is the eigenvalue of the Laplacian on a hyperbolic space. As in AS we are correcting for compensation by including a factor of $1/(1 + (k_c/k)^2)$ which gives the correct energy momentum behaviour on large scales.

The evolution equations for the Green's functions are integrated numerically. In the CDM case we have checked that they match, for $\Omega = 1$, the analytic fit of [9] and, for variable $\Omega$ and $h$ and no sources, the pure CDM transfer function of [10]. For BDM, following [11], we have solved the fluid equations in the tightly coupled regime until this approximation breaks down. We then calculate explicitly the effect due to damping during recombination and finally match onto post-recombination solutions. The "Sachs-Wolfe" term is simplified by discarding the traceless part of the metric, and we model the evolution of the (standard recombination) ionization fraction accurately by integrating the equations derived in [12].

We can fit our results with the function:

$$P(k) = \frac{k\alpha_1^2(G\mu)^2}{1 + (\alpha_2 k) + (\alpha_3 k)^2 + (\alpha_4 k)^3} \left(\frac{1}{1 + (\alpha_5/k)^2}\right)^2 \qquad (4)$$

where the $\alpha$'s can be read off from the table (k in units of $h^2 Mpc^{-1}$).

|  | $\alpha_1$ | $\alpha_2$ | $\alpha_3$ | $\alpha_4$ | $\alpha_5$ | $bG\mu_6$ |
|---|---|---|---|---|---|---|
| $\Omega_{BDM} = .2$ | 143.7 | 23 | 21 | 17.5 | .00031 | 9.6 |
| $\Omega_{BDM} = .4$ | 190.0 | 11.5 | 11 | 12.55 | .0004 | 4.6 |
| $\Omega_{CDM} = .1$ | 195.2 | 47 | 44 | 15.5 | .0002 | 9.9 |
| $\Omega_{CDM} = .3$ | 168.1 | 15.7 | 14.7 | 5.17 | .00037 | 4.3 |

TABLE I. Parameters of four models normalized, to fit the 1 in 6 QDOT IRAS power spectrum; $b$ is the bias factor and $G\mu$ is in units of $10^{-6}$.

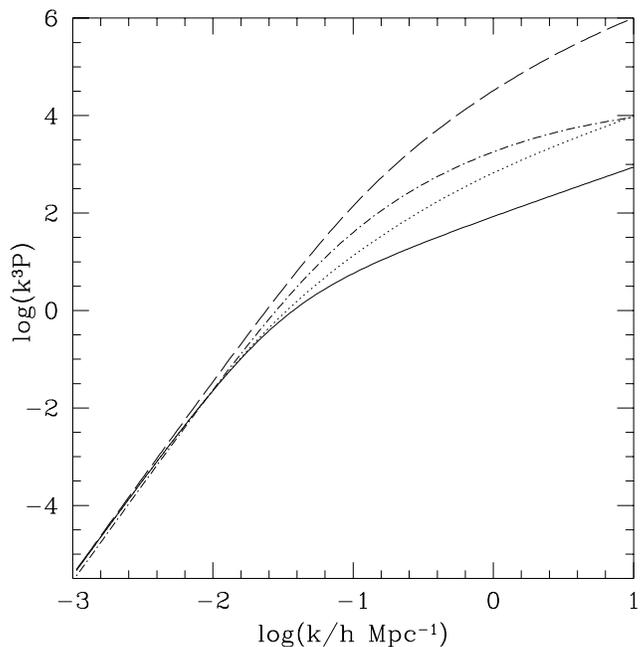

FIG. 1. A plot of the mass fluctuation power spectrum for 4 different models: solid line is strings and BDM ($\Omega = .2$), dotted line is strings and CDM ($\Omega = .2$), long dashed line is strings and CDM ($\Omega = 1$) and dot-dashed line is adiabatic CDM ($\Omega = 1$). They are normalized to agree on large scales.

As expected, by lowering $\Omega$ one can shift the turnover in the power spectra to large enough scales. In fact for the CDM models we confirm [4]: one can obtain the lower $\Omega$ models by simply rescaling, $k \to k/\Omega$, a consequence of how late time evolution of the network is unimportant on the scales we are interested. In Fig.1 we compare two models with a standard CDM model and an $\Omega = 1$ strings and CDM model. Note that, unlike in adiabatic BDM models, in the strings and BDM model there is no remnant of acoustic oscillations, seeing as one is incoherently summing over perturbations seeded at different times.

Although it would be premature to use large scale surveys to make definitive statements about our models we have found it instructive to compare four models with the power spectrum derived in [13] from the 1-in-6 QDOT IRAS redshift survey (Figs.2 and 3). Clearly an $\Omega_{CDM} = .3$, $h = .5$ CDM model continues to have too much power on smaller scales ($k > 0.07$) scales compared to large scales (we have chosen the model in which the turnover best agrees with the data), while in an $\Omega_{CDM} = .1$ model, although having less power on small scales, has a turnover on too large a scale. In a BDM model, the power spectrum agrees well with the data for $k > .04$, but we find that the model that best fits the data on these scales ($\Omega_{BDM} = .2$) has a turnover on too large a scale. A better fit in this sense is then $\Omega_{BDM} = .4$ [14]. In Table 1. we present an estimate of $G\mu$ which comes from normalizing the power spectrum to these scales. Seeing as we are working in linear theory we have opted *not* to normalize to $\sigma_8$ where gravitational collapse has most likely already begun.

Within this model we are unable to make any precise predictions about the cosmic microwave background; flat spacetime simulations indicate that all three modes (scalar, vector and tensor) contribute on equal footing to the final anisotropy a fact which makes it necessary to have a much deeper understanding of the network and its correlation functions. In [4] it was pointed out that low multipole moments of the temperature correlation function are suppressed; for a CDM, $\Omega = .2$ defect-seeded universe, the spectrum for $l < 5$ is lower than for a flat universe. One would expect, however, that for higher multipoles the spectrum is roughly the same in both open and flat models. To reconcile the values of $G\mu$ presented in the table with the one presented in [15], one would need values of $b \simeq 2.5 - 5$.

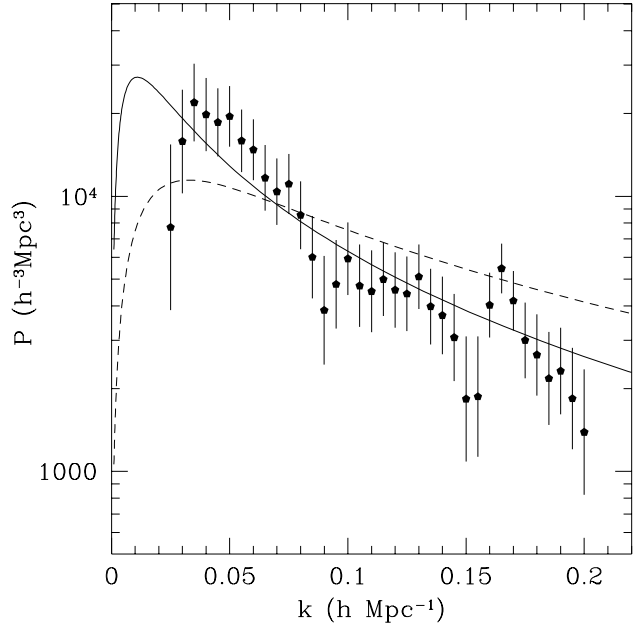

FIG. 2. The power spectrum of two CDM models ($\Omega = .1$ solid line and $\Omega = .3$ dashed line) plotted against the 1 in 6 QDOT IRAS power spectrum.

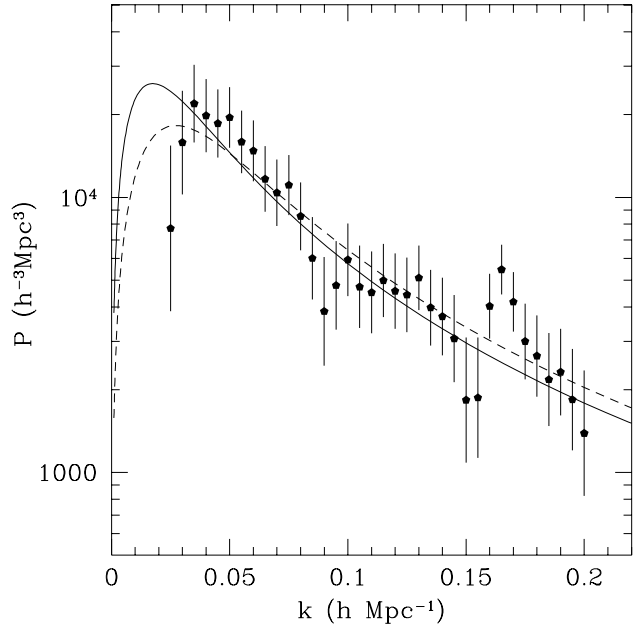

FIG. 3. The power spectrum of two BDM models ($\Omega = .2$ solid line and $\Omega = .4$ dashed line) plotted against the 1 in 6 QDOT IRAS power spectrum.

The evolution of small wavelength perturbations in BDM is qualitatively different from that of perturbations in CDM; collisional (or Silk) damping at around the time of recombination will suppress baryonic perturbations generated before then (unlike in CDM where there is no such effect). However the constant sourcing by the string network will ensure that perturbations are generated on small scales at late times (primordial isocurvature perturbations, i.e. perturbations in the radiation field generated at the phase transition, are on sufficiently small scales to be ignored). Unlike in HDM, the damping becomes irrelevant in a finite time so the tail reflects the $k^{-2}$ behaviour of the structure function on small scales.

This essential difference between the two models may lead to different features in the matter distribution. On small scales in CDM, wakes are competing with an r.m.s. $\delta m/m$ which has been growing for a long time, beginning in the radiation era. The result is that they are not so likely to stand out in a final realization of the matter distribution. In BDM however, small scale perturbations have been wiped out, and therefore it is post-recombination wakes that are responsible for perturbations on small scales. A possible picture is that small scale structures will form in such away as to be organized along large scale coherent structures, a realization of the the "radical biasing" suggested in [2].

This has bearing on another phenomena of some importance; it has been argued that it is likely that a defect-seeded cosmology will be reionized due to early formation of non-linear objects, intimately related to the non-gaussian nature of the perturbations. This has been verified for texture cosmogonies (where approximately 1% of the matter had formed bound objects before a redshift of 50). It is not so clear that one can apply this sort of reasoning to string seeded perturbations; the string network tends to be much denser than other defects and analysis of both CMBR maps [15] and wake prominence in CDM models indicates that string perturbations tend to be much more "gaussian". One expects therefore that reionization is less likely in these (CDM) models [16]. There are additional complications in the BDM models where on the one hand structure formation can be delayed if the medium is ionized but on the other hand wakes can stand out on small scales leading possibly to early star formation. We intend to investigate the consequences of late time recombination in string models in some detail.

In summary, we have looked at string seeded perturbations in open universes and selected a few models that are consistent with current large scale surveys; we have considered models with either baryonic or non-baryonic dark matter and have tried to identify some of the differences in how structures would form on small scales.

## ACKNOWLEDGEMENTS

The author would like to thank A. Albrecht, J. Borrill, D. Coulson, and J.C.R. Magueijo for useful comments and Hume Feldman for generously providing the data in Figs.2, 3. This work was supported by Programa Ciencia (Portugal).